\documentstyle [12pt, epsfig]{article}
\begin{document}

\hfill {CUMQ/HEP 127}

\hfill {\today}

\vskip 0.5in   \baselineskip 24pt

{\Large  \bigskip
      \centerline{\Large Chargino contributions to
                         $\epsilon^{\prime}/ \epsilon$
                  in the left-right supersymmetric model} }

\vskip .6in
\def\bar{\overline}

\centerline{Mariana Frank \footnote{Email: mfrank@vax2.concordia.ca}
and Shuquan Nie \footnote{Email: sxnie@alcor.concordia.ca}}
\bigskip
\centerline {\it Department of Physics, Concordia University, 1455 De
Maisonneuve Blvd. W.}
\centerline {\it Montreal, Quebec, Canada, H3G 1M8}

\vskip 0.5in

{\narrower\narrower We analyze the chargino contributions to the CP
violating ratio $\epsilon^{\prime}/\epsilon$ in the left-right 
supersymmetric model. We study the possibility that these 
contributions alone can saturate the
experimental value of $\epsilon^{\prime}/\epsilon$. We derive 
conservative bounds on supersymmetric flavor violation
parameters in the up squark LL, RR, LR and RL sectors, using
the mass insertion approximation. While the LL
bounds are found to be consistent with the MSSM values, the LR 
constraints are new
and much stronger.}

      PACS number(s): 11.30.Er, 12.60.Jv, 13.25.Es, 14.40.Aq

\newpage

\section{Introduction}

CP violation is still a basic problem to answer in particle theory 
and a good probe of new physics beyond
the Standard Model (SM).
CP violation was first discovered in
the kaon system \cite{kaon}, and recently confirmed in
the B-system \cite{babar, belle}, where assumptions about the gauge 
structure would be verified.
There are two CP violation parameters in the kaon system: the 
indirect CP violation
parameter $\epsilon$, which follows from the mass eigenstates of 
$K^0$ and $\bar{K}^0$ and is given by
the imaginary parts of the diagrams leading to the mass difference 
$\Delta M_{K}$, and the direct CP violation
parameter $\epsilon^{\prime}$, which describes the decay of $K 
\rightarrow 2 ~ \pi$.

The direct CP violation had been confirmed by \cite {cp}, the world 
average value for the CP
violating ratio $\epsilon^{\prime}/\epsilon$ is \cite{pdg}
\begin{equation}
\mathrm{Re} (\epsilon^{\prime} / \epsilon ) =1.8 \pm 0.4 \times 10^{-3}.
\end{equation}
Even there are large theoretical uncertainties and experimental 
errors associated with it ,
$\epsilon^{\prime}/\epsilon$ can put stringent constraints on extensions of the
SM, as for instance general supersymmetric models \cite{susy}, models with
anomalous gauge couplings \cite {agc}, four-generation models 
\cite{four} and models
with additional fermions and gauge bosons \cite {afag}.

In Ref. \cite{fn} we studied $\Delta S=2$ processes in the kaon system
and evaluated $\Delta M_K$ and $\epsilon$ in the left-right 
supersymmetric model
(LRSUSY) as a scenario for new physics.
In this paper we extend the analysis to $\Delta S=1$ processes and 
put more constraints on the LRSUSY
parameter space and CP violation by calculating $\epsilon^{\prime}/\epsilon$.

There are two kinds of diagrams leading to $\Delta S=1$ processes: 
box and penguin diagrams.
(As opposed to $\Delta S=2$ processes, which are mediated by box
diagrams alone.)
The exchange particles in
general supersymmetric models can be gluinos, neutralinos and charginos.
The gluino contributions have been studied widely in literature \cite{gluino},
while chargino contributions have not
received much attention, mainly due to its strong model-dependence 
\cite{chargino}.
The gluino
and neutralino contributions put constraints on squark mixings in the
down sector; there
gluino contributions are dominant and therefore the bounds are
sensitive to the QCD sector in the model.
Studying the chargino contributions would constrain the mixing in the
up squark sector,
independent of the down sector, and thus test models with different
electroweak symmetries
than the MSSM, in  particular the LRSUSY. This is the goal of the 
present article.

Our paper is organized as follows: in section II, after a short
description of the model, we give complete analytical results
for the chargino contributions to $\epsilon^{\prime}/ \epsilon $.
We follow by presenting numerical analysis in section III.
We reach our conclusions in section IV, and in the
appendix we give a summary of the chargino mixing in the LRSUSY, as well as
the loop and vertex functions used, for self-sufficiency of the paper.

\section{The analytic formulas}

\subsection{Effective Hamiltonian for $\Delta S=1$ processes in the LRSUSY}

The left-right extension to the supersymmetric standard model
is based on the gauge group
$SU(3)_C \times SU(2)_L \times SU(2)_R \times U(1)_{B-L}$ 
\cite{lrsusy}. The model has
chiral superfields in left and right handed doublets. The Higgs
sector consists of two bidoublet and four triplet Higgs superfields.
The bidoublet Higgs superfields exist in all versions of LRSUSY and
break the symmetry group to $SU(2)_L \times U(1)_{Y}$. Additional
Higgs representations needed to break $SU(2)_R$ symmetry are commonly
chosen to be triplets which support the
seesaw mechanism. One needs to double the number of Higgs fields
with respect to the non-supersymmetric version to ensure anomaly
cancellations in the fermionic sector. The
superpotential involving these superfields is
\begin{eqnarray}
\label{superpotential}
W & = & {\bf Y}_{Q}^{(i)} Q^T\Phi_{i}i \tau_{2}Q^{c} + {\bf Y}_{L}^{(i)}
L^T \Phi_{i}i \tau_{2}L^{c} + i({\bf Y}_{LR}L^T\tau_{2} \Delta_L L +
{\bf Y}_{LR}L^{cT}\tau_{2}
\Delta_R L^{c}) \nonumber \\
& & + \mu_{LR}\left [Tr (\Delta_L  \delta_L +\Delta_R
\delta_R)\right] + \mu_{ij}Tr(i\tau_{2}\Phi^{T}_{i} i\tau_{2} \Phi_{j}).
\end{eqnarray}
In addition, flavor and CP violating effects arise from the the soft
supersymmetry breaking terms
\begin{eqnarray}
\label{eq:soft}
{\cal L}_{soft}&=&\left[ {\bf A}_{Q}^{i}{\bf Y}_{Q}^{(i)}{\tilde Q}^T\Phi_{i}
i\tau_{2}{\tilde Q}^{c}+ {\bf A}_{L}^{i}{\bf Y}_{L}^{(i)}{\tilde L}^T \Phi_{i}
i\tau_{2}{\tilde L}^{c} + i{\bf A}_{LR} {\bf Y}_{LR}({\tilde L}^T\tau_{2}
\Delta_L{\tilde  L} + L^{cT}\tau_{2} \Delta_R{\tilde L}^{c})
\right.
\nonumber \\
   &+ &\left.{ m}_{\Phi}^{(ij) 2}
\Phi_i^{\dagger}  \Phi_j \right] + \left[( m_{Q_L}^2)_{ij}{\tilde
Q}_{Li}^{\dagger}{\tilde
Q}_{Lj}+ (m_{Q_R}^2)_{ij}{\tilde Q}_{Ri}^{\dagger}{\tilde Q}_{Rj}
\right]+ \left[( m_{L}^2)_{ij}{\tilde
l}_{Li}^{\dagger}{\tilde
l}_{Lj}+ (m_{R}^2)_{ij}{\tilde l}_{Ri}^{\dagger}{\tilde l}_{Rj}
\right]  \nonumber \\
&-& M_{LR}^2 \left[
Tr(  \Delta_R  \delta_R)+ Tr(  \Delta_L
     \delta_L) + h.c.\right]
- [B \mu_{ij} \Phi_{i} \Phi_{j}+h.c.].
\end{eqnarray}

We parameterize all the unknown soft breaking
parameters coming
mostly from the scalar mass matrices using the mass insertion approximation
\cite{MI}. We choose a basis for fermion and sfermion states
in which all the couplings of these particles to neutral gauginos are flavor
diagonal and parameterize flavor changes in
the non-diagonal squark propagators through mixing parameters
$(\delta^q_{ij})_{AB}$, where $i,j=1, 2, 3$ and $A,B=L,R$. The 
dimensionless flavor
mixing parameters used are
\begin{eqnarray}
\label{massins}
(\delta^{q}_{ij})_{LL}&=&\frac{(m^2_{\tilde{q},ij})_{LL}}{m_{\tilde q}^2},~~~
(\delta^{q}_{ij})_{RR}=\frac{(m^2_{\tilde{q},ij})_{RR}}{m_{\tilde
q}^2}, \nonumber \\
(\delta^{q}_{ij})_{LR}&=&\frac{(m^2_{\tilde{q},ij})_{LR}}{m_{\tilde q}^2},~~~
(\delta^{q}_{ij})_{RL}=\frac{(m^2_{\tilde{q},ij})_{RL}}{m_{\tilde q}^2},
\end{eqnarray}
where $m_{\tilde q}^2$ is the average squark mass and
$(m^2_{\tilde{q},ij})_{AB}$ are the
off-diagonal elements which mix squark flavors for both left- and
right- handed squarks with $q=u,d$.

The contributions to $\Delta S=1$ processes
are given by the effective Hamiltonian
\begin{equation}
{\cal H}_{eff}^{\Delta S=1} =\sum_i [ C_i(\mu) Q_i(\mu) +
{\tilde C}_i(\mu) {\tilde Q}_i(\mu)],
\end{equation}
where the relevant operators entering the sum are
\begin{eqnarray}
Q_{3}&=&{\bar d}_L^\alpha \gamma_\mu s_L^\alpha \sum_{q=u,d,s}{\bar
q}_L^\beta \gamma^\mu q_L^\beta,
\nonumber \\
Q_{4}&=&{\bar d}_L^\alpha \gamma_\mu s_L^\beta \sum_{q=u,d,s}{\bar
q}_L^\beta \gamma^\mu q_L^\alpha,
\nonumber \\
Q_{5}&=&{\bar d}_L^\alpha \gamma_\mu s_L^\alpha \sum_{q=u,d,s}{\bar
q}_R^\beta \gamma^\mu q_R^\beta,
\nonumber \\
Q_{6}&=&{\bar d}_L^\alpha \gamma_\mu s_L^\beta \sum_{q=u,d,s}{\bar
q}_R^\beta \gamma^\mu q_R^\alpha,
\nonumber \\
Q_{7}&=&\frac{Q_d e}{8 \pi^2} m_s {\bar d}_L^\alpha \sigma^{\mu \nu}
s_R^\alpha F_{\mu \nu},
\nonumber \\
Q_{8}&=&\frac{g}{8 \pi^2} m_s {\bar d}_L^\alpha \sigma^{\mu \nu} 
t^a_{\alpha \beta}
s_R^\beta G_{\mu \nu}^a,
\end{eqnarray}
where $q_{R,L}=P_{R,L} q$ with $P_{R,L} =(1 \pm \gamma_5)/2$, and
$\alpha,~ \beta$ are color indices. The operators $\tilde{Q}_i$ are obtained
from $Q_i$ by the exchange $L \leftrightarrow R$.
Because of the left-right symmetry,
we must consider all contributions from both chirality operators.
We neglect the higgsino
couplings (proportional to $m_s$ or $m_d$) and include single mass
insertion only. To this order, the contributions are
\begin{figure}
\centerline{ \epsfysize 5.5 in
\rotatebox{360}{\epsfbox{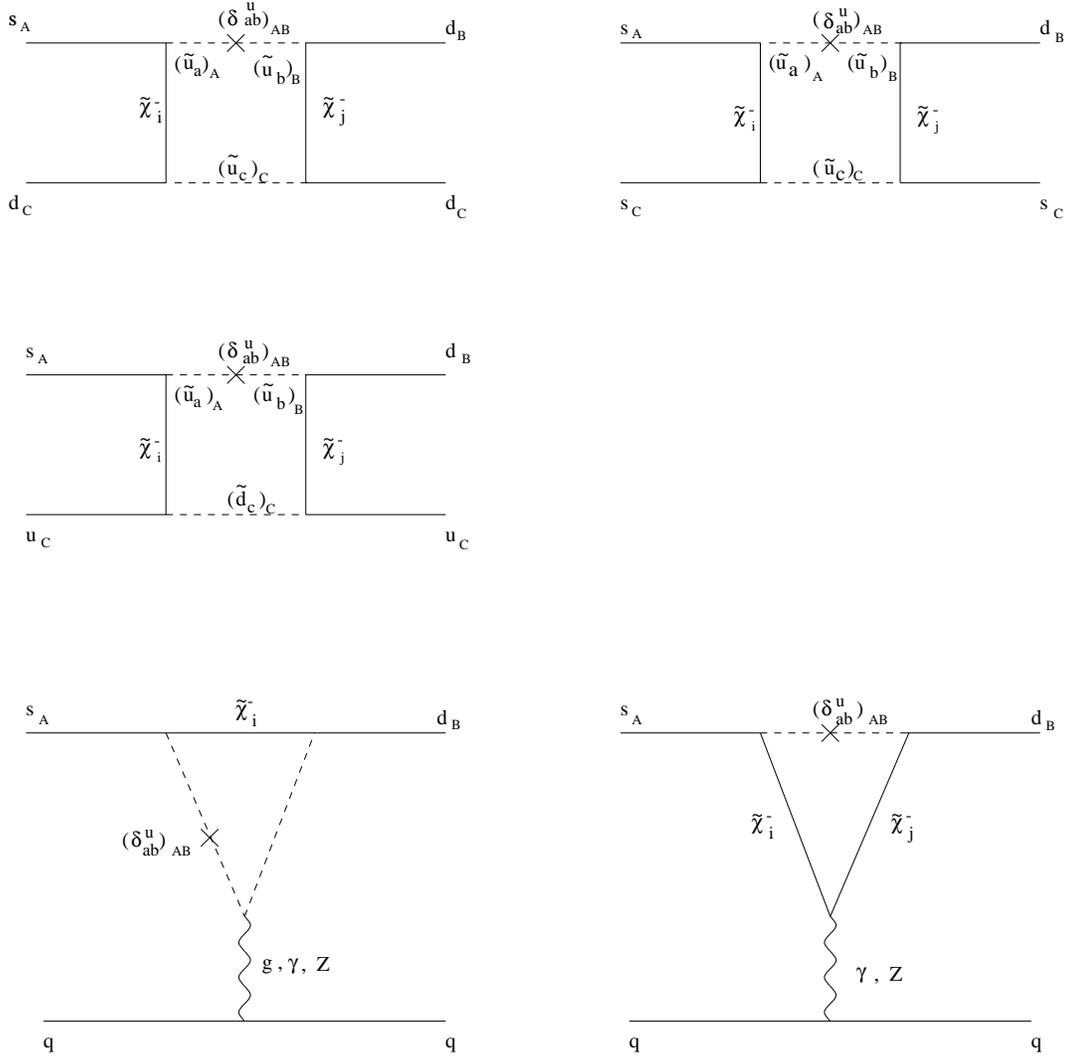}}  }
\caption{ Leading supersymmetric box and penguin diagrams
contributing to $\epsilon^{\prime}$, $A, ~B, ~C=(L,~R)$.}
\protect \label{epsilonprime}
\end{figure}
\begin{equation}
C_i=C_i^{box}+C_i^{g-penguin}+C_i^{\gamma-penguin}+C_i^{Z-penguin},
\end{equation}
with
\begin{eqnarray}
\label{equation8}
C_4^{box}&=&\frac{\alpha_W^2}{m_{\tilde{q}}^2} \sum_{i,j=1}^4 
\sum_{a,b=1}^3 |V_{i1}|^2 |V_{j1}|^2  K^*_{a2} K_{b1}
B^{(1)}_{\tilde{\chi}^-} (x_{\tilde{\chi}_i^- \tilde{q}},
x_{\tilde{\chi}_j^- \tilde{q}}) (\delta^u_{ab})_{LL},
\nonumber \\
C_5^{box}&=&\frac{\alpha_W^2}{2m_{\tilde{q}}^2}  \sum_{i,j=1}^4 
\sum_{a,b=1}^3  |V_{i1}|^2 |V_{j1}|^2  K^*_{a2} K_{b1}
B^{(2)}_{\tilde{\chi}^-} (x_{\tilde{\chi}_i^- \tilde{q}},
x_{\tilde{\chi}_j^- \tilde{q}}) (\delta^u_{ab})_{LL},
\nonumber \\
C_3^{g-penguin}&=&-\frac{\alpha_s \alpha_W}{ 3m_{\tilde{q}}^2} 
\sum_{i=1}^4 \sum_{a,b=1}^3  | V_{i1}|^2  K_{a2}^* K_{b1}
f_g (x_{\tilde{\chi}_i^- \tilde{q}}) (\delta_{ab}^u)_{LL},
\nonumber \\
C_4^{g-penguin}&=&\frac{\alpha_s \alpha_W}{ m_{\tilde{q}}^2} 
\sum_{i=1}^4 \sum_{a,b=1}^3  | V_{i1}|^2  K_{a2}^* K_{b1}
f_g (x_{\tilde{\chi}_i^- \tilde{q}}) (\delta_{ab}^u)_{LL},
\nonumber \\
C_5^{g-penguin}&=&-\frac{\alpha_s \alpha_W}{ 3m_{\tilde{q}}^2} 
\sum_{i=1}^4 \sum_{a,b=1}^3  | V_{i1}|^2  K_{a2}^* K_{b1}
f_g (x_{\tilde{\chi}_i^- \tilde{q}}) (\delta_{ab}^u)_{LL},
\nonumber \\
C_6^{g-penguin}&=&\frac{\alpha_s \alpha_W}{ m_{\tilde{q}}^2} 
\sum_{i=1}^4 \sum_{a,b=1}^3  | V_{i1}|^2  K_{a2}^* K_{b1}
f_g (x_{\tilde{\chi}_i^- \tilde{q}}) (\delta_{ab}^u)_{LL},
\nonumber \\
C_8^{g-penguin}&=&-\frac{\pi \alpha_W }{ m_{\tilde{q}}^2}
\sum_{i=1}^4 \sum_{a,b=1}^3 K_{a2}^* K_{b1}  \{ | V_{i1}|^2
F_2 (x_{\tilde{\chi}_i^- \tilde{q}}) (\delta_{ab}^u)_{LL}
\nonumber \\
& & + U_{i1} V_{i1}^* \frac{m_{\tilde{\chi}_i^-}}{m_s} F_4 
(x_{\tilde{\chi}_i^- \tilde{q}})
(\delta_{ab}^u)_{LR}  \},
\nonumber \\
C_7^{\gamma-penguin}&=&-\frac{\pi \alpha_W }{ m_{\tilde{q}}^2}
\sum_{i=1}^4 \sum_{a,b=1}^3 K_{a2}^* K_{b1}  \{ | V_{i1}|^2 \left [
F_1 (x_{\tilde{\chi}_i^- \tilde{q}})+ Q_u F_2 (x_{\tilde{\chi}_i^- 
\tilde{q}}) \right ] (\delta_{ab}^u)_{LL}
\nonumber \\
& & + U_{i1} V_{i1}^* \frac{m_{\tilde{\chi}_i^-}}{m_s}
\left [ F_3 (x_{\tilde{\chi}_i^- \tilde{q}}) + Q_u F_4 
(x_{\tilde{\chi}_i^- \tilde{q}}) \right ]
(\delta_{ab}^u)_{LR}  \},
\nonumber \\
C_3^{Z-penguin}&=&\frac{\alpha_W^2}{ m_Z^2 \cos^2 \theta_W} 
\sum_{a,b=1}^3 K_{a2}^* K_{b1}  \{ \sum_{i=1}^4 | V_{i1}|^2
(-1+\frac{4}{3} \sin^2 \theta_W) f_Z (x_{\tilde{\chi}_i^- \tilde{q}})
\\
& & +\sum_{i, j=1}^4 V_{i1} V_{j1}^*
  \left ( (Z_L)^R_{ij} f_Z^{(1)} (x_{\tilde{\chi}_i^- 
\tilde{q}},x_{\tilde{\chi}_j^- \tilde{q}})
+(Z_L)^L_{ij} f_Z^{(2)} (x_{\tilde{\chi}_i^- 
\tilde{q}},x_{\tilde{\chi}_j^- \tilde{q}}) \right )
  \} (\delta_{ab}^u)_{LL},
\nonumber
\end{eqnarray}
where $x_{ab}=m^2_a/m^2_b$. There is no chargino box contribution to 
$C_3$ and $C_6$ because of
the color structure. Similarly, there is no photon and $Z_L$ penguin 
contribution to $C_4$ and $C_6$.
The photon penguin gives zero contribution to $C_3$ and $C_5$ after 
the sum over quarks.
Similarly, the $Z_L$ penguin gives zero contribution to $C_5$.
We neglect contributions from the $Z_R$ penguin since they are
smaller by $m^2_{Z_L}/m^2_{Z_R}$ than the contributions from the
$Z_L$ penguin.
The notations of various vertices, mixing
matrices and functions are defined in the appendix.
The coefficients $\tilde {C}_i$ are obtained from
the coefficients $C_i$ by the exchange  $L \leftrightarrow R$.

\subsection{Hadronic Matrix Elements}
We list here for completeness the relevant matrix elements of operators,
which can be found in Ref. \cite{elements}
\begin{eqnarray}
\left < (\pi \pi)_{I=0} | Q_3 | K \right > &=& \frac{X}{3},
\nonumber \\
\left < (\pi \pi)_{I=0} | Q_4 | K \right > &=& X,
\nonumber \\
\left < (\pi \pi)_{I=0} | Q_5 | K \right > &=&- \frac{Y}{3},
\nonumber \\
\left < (\pi \pi)_{I=0} | Q_6 | K \right > &=& -Y,
\nonumber \\
\left < (\pi \pi)_{I=0} | Q_8 | K \right > &=&
-\sqrt{\frac{3}{2}} \frac{1}{16 \pi^2} \frac{11}{2} \frac{m_s}{m_s+m_d}
\frac{f_K^2}{f_{\pi}^3}m_K^2 m_{\pi}^2,
\nonumber \\
\left < (\pi \pi)_{I=2} | Q_i | K \right > &=& 0,~~i=3 \cdots 6,8
\end{eqnarray}
where
\begin{eqnarray}
X &=& \sqrt{\frac{3}{2}} f_{\pi} \left ( m_K^2 - m_{\pi}^2 \right ),
\nonumber \\
Y &=& \sqrt{\frac{3}{2}} (f_K -f_{\pi}) \left ( \frac{m_K^2}{m_s+m_d} 
\right )^2.
\end{eqnarray}
 From Ref. \cite{Q7}, $\left < (\pi \pi)_{I=0,2} | Q_7 | K \right >$ 
is proportional to the photon condensate,
thus it must be very small and negligible. Accordingly the contribution to
$\epsilon^{\prime}$ of $C_7^{\gamma-penguin}$ can be neglected.
The matrix elements of the
operators ${\tilde Q}_i$ can be obtained by multiplying the
corresponding matrix elements of $Q_i$ by $(-1)$.

Putting all the above together, we can calculate the CP violating ratio
$\epsilon^{\prime}/\epsilon$
\begin{eqnarray}
\frac{\epsilon^{\prime}}{\epsilon} &=& - \frac{\omega}{\sqrt{2} 
|\epsilon| \mathrm{Re}A_0}
\left ( \mathrm{Im} A_0 -\frac{1}{\omega} \mathrm{Im} A_2 \right ),
\end{eqnarray}
where $\omega=\mathrm{Re}A_2/\mathrm{Re}A_0$ and the amplitudes are defined as
\begin{equation}
A_I e^{i \delta_I} = < (\pi \pi)_I | {\cal H}_{eff}^{\Delta S=1} | K >,
\end{equation}
with the isospin of the final two-pion state $I=0, ~2$, and 
$\delta_I$'s are strong phases
induced by final state interactions, $\delta_2-\delta_0$ is close to $\pi/4$.

\section{Numerical Analysis}

In this section we present the results of our analysis for the
individual bounds on $(\delta^u_{12})_{AB}$, obtained by
selecting only one
source of flavor violation and neglecting interference between
different sources. Setting the CKM phase to zero, therefore there is 
no SM contribution to $\epsilon^{\prime}$.
The constraints are obtained by requiring
the LRSUSY contributions alone to saturate the experimental value for
$\epsilon^{\prime}/ \epsilon$. Therefore the bounds are conservative.

We choose all trilinear scalar couplings in the soft supersymmetry
breaking Lagrangian
to be universal: $A_{ij}=A \delta_{ij}$ and
$\mu_{ij}=\mu \delta_{ij}$, and we fix $A$ to be $100$ GeV, the
higgsino mixing parameter $\mu=200$ GeV, and $\tan \beta=5$
throughout the analysis. We also demand that the chargino masses satisfy
the current experimental bounds.

The bounds on $(\delta_{12}^u)_{LL}$ are presented in Table {\bf 1}.
 From Eq. \ref{equation8}, the box, gluon-penguin and photon-penguin 
contributions are proportional
to $1/m_{\tilde{q}}^2$, while the $Z_L$-penguin contribution is 
proportional to $1/m_Z^2$. Thus the $Z_L$-penguin
contribution dominates over all other contributions as 
$m_{\tilde{q}}^2$ are expected to
be larger than $m_Z^2$.
Our bounds on $(\delta_{12}^u)_{LL}$ are comparable with previous 
bounds of Khalil and Lebedev
\cite{chargino},
and they are also the same order of magnitude as the bounds on 
$(\delta_{12}^d)_{LL}$ obtained from the
gluino contributions to $\epsilon^{\prime}$ \cite{ggms}.
The values we obtained are always of ${\cal
O}(10^{-1})$ and fairly stable over a large
range of chargino and squark mass parameters. What is different in
this model from the MSSM is that, due to the mass
parameters and symmetry of the model in the chargino sector, to a
good approximation we get the same bounds on
$(\delta_{12}^u)_{RR}$.

{\center
\begin{tabular}{|c|c|c|c|c|}  \hline
$M_L $ $\slash$ $ m_{\tilde{q}}$ (GeV) & 300  & 500 & 700 & 900\\
\hline
\hline
      150 &$0.10$

          &$0.14$

          &$0.16$

          &$0.18$\\
      250 &$0.18 $

          &$0.29 $

          &$0.42 $

          &$0.57 $\\
      350 &$0.23 $

          &$0.30 $

          &$0.38 $

          &$0.46 $\\
      450 &$0.29 $

          &$0.38 $

          &$0.42 $

          &$0.50 $\\
\hline
\end{tabular}

\vspace{0.5cm}
${\bf  Table ~~1} ~~$ Limits on $|\mathrm{Im} (\delta_{12}^u)_{LL}|$
from $\epsilon^{\prime}$ for different values of
$M_L=M_R$ and $m_{\tilde{q}}$,
with $\tan \beta=5$ and $\mu=200$ GeV. The bounds
are insensitive to $\tan \beta$ in the range of $2-30$ and $\mu$
in the range of $200-500$ GeV.
}

The bounds on $(\delta_{12}^u)_{LR}$ are presented in Table {\bf 2}.
These bounds come from
the gluon and photon penguins; as before we would obtain
approximately the same bounds on
$(\delta_{12}^u)_{LR}$ as on $(\delta_{12}^u)_{RL}$.
As there is a large factor
$m_{\tilde{\chi}_i^-}/m_s$ in the penguin LR contributions, the 
bounds on $(\delta_{12}^u)_{LR}$ are
tighter by 1-2 orders of magnitude than
the bounds on  $(\delta_{12}^u)_{LL}$.
These bounds are also less
stable than the previous ones, and can vary
by a factor of $10^2$ over the range of chargino and squark masses
explored. Both sets of bounds are rather
insensitive to other parameters, such as $\tan \beta$ or the higgsino
mixing parameter $\mu$, over a
low-intermediate range of values.

{\center
\begin{tabular}{|c|c|c|c|c|}  \hline
$M_L $ $\slash$ $ m_{\tilde{q}}$ (GeV) & 300  & 500 & 700 & 900\\
\hline
\hline
      150 &$4.48 \times 10^{-4} $

          &$3.84 \times 10^{-3} $

          &$1.76 \times 10^{-2} $

          &$5.79 \times 10^{-2} $\\
      250 &$1.40 \times 10^{-3} $

          &$5.55 \times 10^{-3} $

          &$2.02 \times 10^{-2} $

          &$5.74 \times 10^{-2} $\\
      350 &$1.34 \times 10^{-3} $

          &$5.08 \times 10^{-3} $

          &$1.58 \times 10^{-2} $

          &$4.09 \times 10^{-2} $\\
      450 &$1.12 \times 10^{-3} $

          &$2.16 \times 10^{-3} $

          &$1.21 \times 10^{-2} $

          &$2.98 \times 10^{-2} $\\
\hline
\end{tabular}

\vspace{0.5cm}
${\bf  Table ~~2} ~~$ Limits on $|\mathrm{Im} (\delta_{12}^u)_{LR}|$
from $\epsilon^{\prime}$ for different values of
$M_L=M_R$ and $m_{\tilde{q}}$,
with $\tan \beta=5$ and $\mu=200$ GeV. The bounds
are insensitive to $\tan \beta$ in the range of $2-30$ and $\mu$
in the range of $200-500$ GeV.
}

\section{Conclusions}

We have studied the chargino contributions to the CP violating ratio
$\epsilon^{\prime}/\epsilon$ in the LRSUSY.  Assuming the CP
violation to arise from the
supersymmetric contributions only, we derived bounds on the imaginary
parts of the supersymmetric
flavor violation parameters in the LL, RR, LR and RL sectors of the
up squark mass matrix, under
the assumption that only one such insertion dominates.  The bounds in
the LL sector are of
${\cal O}(10^{-1})$, they
agree with the corresponding bounds obtained in the MSSM, and are weak
compared to down squark mass
insertion bounds. The bounds on the LR mass insertions are much
stronger (of ${\cal O}(10^{-4}-10^{-2})$)
and new to this analysis: no similar bounds exist for the MSSM.
Comparing this analysis with the
previous results coming from the up squark sector of the
$K^0-{\bar K}^0$ parameter $\epsilon$ \cite{fn}, we see that the
constraints from
$\epsilon^{\prime}/\epsilon$ are weaker by one or more orders of magnitude.
Therefore, even if more than
one mass insertion would drive the CP violation, the constraints 
cannot be made to
coincide. Thus, in the LRSUSY, as in the MSSM, supersymmetric contributions
coming from complex mass
insertions in the squark mass matrices fail to saturate both
$\epsilon$ and $\epsilon^{\prime}/\epsilon$, and thus cannot be
responsible alone for the CP
violation.

\noindent {\bf Acknowledgements}

This work was funded in part by NSERC of Canada (SAP0105354).

\newpage

\begin{appendix}

\noindent {\Large {\bf Appendix}}
\label{appendix}

For self-sufficiency, we list the mass-squared
matrices for charginos and neutralinos,
relevant Feynman rules and functions used for this calculation.

The terms relevant to the masses of charginos in the Lagrangian are
\begin{equation}
{\cal L}_C=-\frac{1}{2}(\psi^+, \psi^-) \left ( \begin{array}{cc}
                                                           0 & X^T \\
                                                           X & 0
                                                         \end{array}
                                                 \right ) \left (
\begin{array}{c}
                                                                  \psi^+ \\
                                                                  \psi^-
                                                                  \end{array}
                                                           \right ) + H.c. \ ,
\end{equation}
where $\psi^+=(-i \lambda^+_L, -i \lambda^+_R, \tilde{\phi}_{u1}^+,
\tilde{\phi}_{d1}^+)^T$
and $\psi^-=(-i \lambda^-_L, -i \lambda^-_R, \tilde{\phi}_{u2}^-,
\tilde{\phi}_{d2}^-)^T$, and
\begin{equation}
X=\left( \begin{array}{cccc}
                               M_L & 0 & g_L \kappa_u & 0  \\
                               0 & M_R & g_R \kappa_u & 0  \\
                               0 & 0 & 0 & -\mu \\
                               g_L \kappa_d & g_R \kappa_d & -\mu & 0
                  \end{array}
            \right ),
\end{equation}
where we have taken, for simplification, $\mu_{ij}=\mu \delta_{ij}$. 
The chargino mass
eigenstates $\tilde{\chi}_i$ are obtained by
\begin{eqnarray}
\tilde{\chi}_i^+=V_{ij}\psi_j^+, \ \tilde{\chi}_i^-=U_{ij}\psi_j^-, \
i,j=1, \ldots 4,
\end{eqnarray}
with $V$ and $U$ unitary matrices satisfying
\begin{equation}
U^* X V^{-1} = M_D,
\end{equation}
The diagonalizing matrices $U^*$ and $V$ are obtained by
computing the eigenvectors corresponding
to the eigenvalues of $X^{\dagger} X$ and $X X^{\dagger}$, respectively.

The vertices of $Z_L$-chargino-chargino are given by
\begin{eqnarray}
(Z_L)^L_{ij} &=& V_{i1} V_{j1}^*+\frac{1}{2} 
V_{i3}V_{j3}^*+\frac{1}{2} V_{i4} V_{j4}^*-\sin^2 \theta_W 
\delta_{ij},
\\
(Z_L)^R_{ij} &=& U_{i1}^* U_{j1}+\frac{1}{2} 
U_{i3}^*U_{j3}+\frac{1}{2} U_{i4}^* U_{j4}-\sin^2 \theta_W 
\delta_{ij}.
\end{eqnarray}

The relevant functions are listed in the following
\begin{eqnarray}
B^{(1)}_{\tilde{\chi}^-} (x,y)&=&\frac{1}{8(x-y)}\left [ \frac{-3 x^2 
+4x-1+2 x^2 \log{x} }{(x-1)^3}
-(x \rightarrow y) \right ],
\\
B^{(2)}_{\tilde{\chi}^-} (x,y)&=&\sqrt{xy} \frac{1}{2(x-y)} \left [ 
\frac{- x^2 +1+2 x \log{x} }{(x-1)^3}
-(x \rightarrow y) \right ],
\\
f_g (x) &=& \frac{1-6 x +18 x^2 -10 x^3 -3 x^4 +12 x^3 \log{x}}{18 (x-1)^5},
\\
f_\gamma (x) &=& \frac{22-60 x +45 x^2 -4 x^3 -3 x^4 +3(3-9 x^2+4 
x^3) \log{x}}{27 (x-1)^5},
\\
F_1 (x) &=& \frac{-1+9 x +9 x^2 -17 x^3 + 18 x^2 \log{x}+6 x^3 
\log{x} }{12 (x-1)^5},
\\
F_2 (x) &=& \frac{-1-9 x +9 x^2 + x^3 - 6 x \log{x}-6 x^2 \log{x} }{6 (x-1)^5},
\\
F_3 (x) &=& \frac{1+4 x -5 x^2 + 4 x \log{x} +2 x^2 \log{x} }{2 (x-1)^4},
\\
F_4 (x) &=& -x^2 \frac{5-4 x - x^2 + 2  \log{x} +4 x \log{x} }{2 (x-1)^4},
\\
f_Z (x) &=& \frac{-1+4 x -3 x^2 +2 x^2 \log{x}}{8 (x-1)^3},
\\
f_Z^{(1)} (x, y) &=& \frac{1}{2(x-y)} [ \frac{x^2 \log{x} }{(x-1)^2} 
- \frac{1}{x-1} - (x \rightarrow y) ],
\\
f_Z^{(2)} (x, y) &=& \sqrt{xy} \frac{1}{(x-y)} [ \frac{-x \log{x} 
}{(x-1)^2} + \frac{1}{x-1} - (x \rightarrow y) ].
\end{eqnarray}

\end{appendix}

\bibliographystyle{unsrt}

\end{document}